\input epsf
\documentstyle[11pt,paspconf]{article}

\def\gs{\mathrel{\raise0.35ex\hbox{$\scriptstyle >$}\kern-0.6em
\lower0.40ex\hbox{{$\scriptstyle \sim$}}}}
\def\ls{\mathrel{\raise0.35ex\hbox{$\scriptstyle <$}\kern-0.6em
\lower0.40ex\hbox{{$\scriptstyle \sim$}}}}
\def\w{\omega(\theta)}

\begin{document}

\title{The Clustering of Faint Galaxies on Small Angular Scales}

\author{Tereasa G. Brainerd and Casey J. Law}
\affil{Boston University, Dept. of Astronomy, Boston, MA 02215}

\author{James Brauher}
\affil{IPAC, California Institute of Technology, Pasadena, CA 91125}

\author{S. G. Djorgovski and Ken Banas}
\affil{California Institute of Technology,
105-24, Pasadena, CA 91125}

\begin{abstract}
We present a preliminary measurement of the angular
clustering of faint ($R \le 25$) field galaxies
in which we concentrate on the behavior of $\w$ on small angular
scales ($\theta \ls 10''$).
The galaxies are strongly clustered
and $\w$ is well-characterized by a power law of the
form $A_\omega \theta^{-\delta}$. 
The best-fitting value of the power law index, $\delta$, is, however,
steeper than the fiducial value of $\delta = 0.8$, indicating
that there are more pairs of galaxies separated by $\theta \ls 10''$ in our
sample than would be otherwise expected.
Using the best-fitting form of $\w$, we estimate that $\sim 10\%$ of the
galaxies are in physically close pairs (separations $\ls 21 h^{-1}$~kpc).
This is a factor of order 2 larger than local galaxy
samples but comparable to galaxy samples with $\left< z \right> \sim 0.4$.
The mean redshift of our galaxies is of order 0.95, and, therefore,
our result suggests
that there was little or no evolution in the merger rate of galaxies between
$z \sim 1$ and $z \sim 0.4$.
\end{abstract}


\section{Introduction}

The angular clustering of faint $R$-selected
field galaxies has been studied extensively
(e.g., Efstathiou et al.\ 1991;
Roche et al.\ 1993, 1996; Brainerd, Smail
\& Mould 1995; Hudon \& Lilly 1996; Lidman \& Peterson
1996; Villumsen, Freudling \& da Costa 1996; Woods \& Fahlman 1997), and a
prime motivation of these studies has been to investigate the nature of
the faint field population.  In particular, it is possible to
infer the effective correlation
length of the sample and the rate at which clustering evolves
from a combination of the amplitude of
the angular autocorrelation function, $\w$, and
the redshift distribution of the faint galaxies, $N(z)$.
These observations can then be used
to link properties of the faint field population with
samples of local galaxies. While the exact interpretation remains
controversial, it is generally accepted that overall
$\w$ is fitted well by a power law of the form
$\theta^{-0.8}$ (although see Infante \& Pritchet (1995) for evidence
of a flattening in the power-law coefficient at faint limits).

Here we investigate the clustering of faint galaxies and focus on the
behavior of $\w$ at small angular separations.
We obtain a clear measurement of $\w$ on scales of $\theta < 10''$ whereas
previous investigations have been
largely limited to scales of $\theta \gs 20''$.  Additionally, we use the
clustering properties of the galaxies
to estimate the number of pairs of galaxies that are physically
close to each other in space (separations of $\ls 21 h^{-1}$~kpc).

\section{Observations}

The data consist of deep $R$-band imaging of
11 independent fields that were obtained in good conditions
with the Low Resolution Imaging Spectrograph on 
the 10-m Keck-I telescope.  Each of the $6' \times 8'$ fields 
is centered on a high
redshift quasar with high galactic latitude; however, the presence
of the quasar in the field is irrelevant to the present investigation
(i.e., the presence of a small group of galaxies at the redshift of
the quasar will not influence the results below).
The galaxy catalogs are complete
to $R = 25.0$ and the apparent magnitudes of
the galaxies have been corrected for extinction.  In order
to reduce the stellar contamination in the object catalogs, only objects
with $R \ge 21$ are considered in the analysis below. 
There is, of course, some residual stellar contamination of the galaxy catalogs
at faint limits and
we estimate that to be:
$\sim$16\% ($21.0 \le R \le 24.0$),
$\sim$13\% ($21.0 \le R \le 24.5$),
$\sim$11\% ($21.0 \le R \le 25.0$).
The integral
constraints vary little from field to field due
to the use of the same detector in all cases as well as the lack of
very large, bright galaxies in the fields.  

\section{Analysis and Results}

To compute the angular clustering of the faint galaxies
we use the Landy \& Szalay (1993) estimator:
\begin{equation}
\w
= \frac{DD - 2DR + RR}{RR}
\end{equation}
where $DD$, $DR$, and $RR$ are the number
of unique data-data, data-random, and random-random pairs within
a given angular separation bin.  Regions of the
frame where faint galaxy detection was either lower than average or
impossible (e.g., due to the presence of bright stars and galaxies)
were masked out when computing $DR$ and $RR$.
Raw correlation functions (uncorrected for
stellar contamination or the integral constraint) were determined for
each of the fields, from which a mean correlation function was
computed.  

The results for the mean raw correlation function are shown
in Figure~1, where the error bars show the standard deviation in the
mean.  From top to bottom, the panels show the results for objects with
$21.0 \le R \le 24.0$, $21.0 \le R \le 24.5$, and $21.0 \le R \le 25.0$,
respectively.  Also shown are the formal best-fitting power 
laws of the form $\theta^{-\delta}$
(solid lines) and the best-fitting power laws of the form $\theta^{-0.8}$
(dashed lines).
The power laws in the figure have been suppressed by the appropriate integral
constraints and no correction for residual
stellar contamination has been applied.

\begin{figure}
\plotfiddle{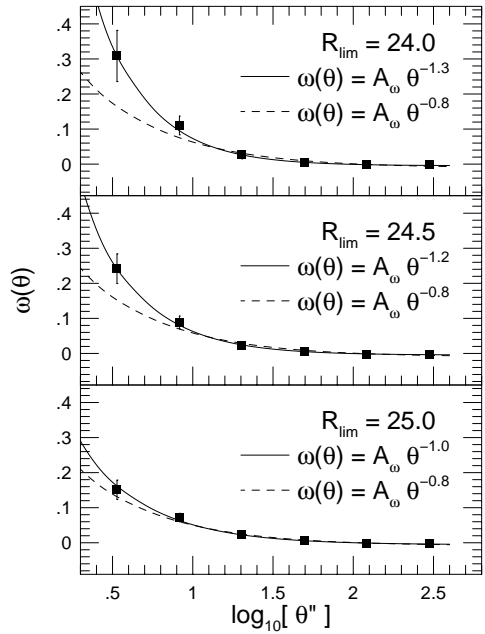}{4.0in}{-90}{35.0}{35.0}{-120.0}{312.0}
\vspace{-2.0cm}
\caption{Mean correlation functions computed using the 11 independent
estimates of $\w$ that were obtained in each magnitude bin.
}
\label{fig-1}
\end{figure}

The number of pairs of galaxies that we observe to be separated
by $\theta \sim 3''$ is larger than the number predicted by the
fiducial $\theta^{-0.8}$ power law (i.e., the power law that
is typically obtained from
measurements that have been performed on scales of $\theta \gs 10''$).  
This is consistent with the results of
Carlberg et al.\ (1994) and Infante et al.\ (1996) who both found
$\w$ to have a higher amplitude on small angular scales ($\theta \ls 6''$)
than a
simple inward extrapolation of $\w$ as measured at large angular scales. 
As yet, however, it is unclear
whether the steepening of $\w$ is due to the existence of a population of
``companion'' galaxies (which are not seen at the present epoch) or
luminosity enhancement (e.g., due to interactions)
of intrinsically faint galaxies that are in pairs.

In the absence of significant luminosity enhancement, 
we can estimate the number of pairs of galaxies that are physically
close to each other simply by using the following probability:
\begin{equation}
P = \int_\beta^\theta 2\pi\rho\alpha \exp(-\pi\rho\alpha^2) \;\; d\alpha
\end{equation}
(e.g., Burkey et al.\ 1994), where $\rho$ is the number density of
galaxies brighter than the faintest member in a pair of galaxies that is
a candidate for close physical
separation, $\theta$ is the observed angular separation between
the galaxies, and $\beta$ is the smallest separation observed between
all detected galaxies ($\beta \sim 1''$ in our data).  Using Eqn.\ (2) 
we compute the number of pairs of galaxies for which $P \le 0.05$ and
$P \le 0.10$ in our data.  Additionally, we use Monte Carlo simulations
(in which the magnitudes of the galaxies are shuffled at random) to
calculate the number of pairs of galaxies that would have $P \le 0.05$
and $P \le 0.10$ simply by chance.  The latter step allows the removal
of random superpositions from the estimate of the ``true'' number of
close pairs in the sample.

Below $\theta \sim 3''$ there are fewer pairs of galaxies with
$P \le 0.05$ and $P \le 0.10$ in the actual data than are expected
in a random distribution (i.e., based on the Monte Carlo simulations), 
indicating that we are undercounting the very closest pairs due to blending
of the images.  Using our measured $\w$, however,
we can correct
the faint pair counts on scales $1'' \le \theta \le 2''$ and estimate
the fraction of galaxies in our sample that are in truly close physical pairs.

Based on a simple extrapolation of the CFRS redshift distribution, 
we expect that the mean redshift of our galaxies is $\sim 0.95$ and, hence,
if $\Omega_0 = 1$, physical pairs of galaxies that are separated by
$\theta \le 5''$ will be within $21 h^{-1}$~kpc of each other.  The
best-fitting power law form of $\w$ (corrected for stellar contamination
and the integral constraint) then yields an estimate of the
pair fraction at this
physical separation of $\sim 10\%$ for $P\le 0.10$, which agrees with
the results obtained by Carlberg et al.\ (1994) for the
fraction of galaxy pairs with separations $\ls 19h^{-1}$~kpc at
$\left< z \right> \sim 0.4$.  This suggests, therefore, that little
evolution in the merger rate of galaxies occurred between 
$z \sim 1$ and $z \sim 0.4$.

\section{Future Work}

A significant amount of work remains to be done on
this project, including a clustering analysis of 7 additional
independent fields and a more rigorous study of the number of
pairs of faint galaxies located at close physical separation.

\acknowledgments
Financial support under NSF contract
AST-9616968 (TGB) and a Boston University Presidential Graduate
Fellowship (CJL)  are gratefully
acknowledged.  The observations were obtained at
the W.\ M.\ Keck Observatory, which is operated jointly by the California
Institute of Technology and the University of California.
Data analysis was performed exclusively on the Origin2000 at
Boston University's Scientific Computing \& Visualisation facility.


\begin{references}
\reference Brainerd, T.G., Smail, I. \& Mould, J.R., 1995, MNRAS, 275, 781
\reference Burkey, J. M., Keel, W. C., Windhorst, R. A., \& Franklin, B. E., 1994,
ApJ, 429, L13
\reference Carlberg, R. G., Pritchet, C. J., \& Infante, L. 1994, ApJ, 435, 540
\reference Efstathiou, G., Bernstein, G., Katz, N., Tyson, J.A., \&
Guhathakurta,
P., 1991, ApJ, 380, L47
\reference Hudon, J.D. \& Lilly, S.J., 1996, ApJ, 469, 519
\reference Infante, L. \& Pritchet, C.J., 1995, ApJ, 439, 565
\reference Infante, L., de Mello, D. \& Menanteau, F. 1996, ApJ, 469, L85
\reference Landy, S.D. \& Szalay, A.S., 1993, ApJ, 412, 64
\reference Lidman, C.E. \& Peterson, B.A., 1996, MNRAS, 279, 1357
\reference Roche, N., Shanks, T., Metcalfe, N., \& Fong, R., 1993, MNRAS, 263,
360
\reference Roche, N., Shanks, T., Metcalfe, N., \&
Fong, R., 1996, MNRAS, 280, 397
\reference Woods, D. \& Fahlman, G.G., 1997, ApJ, 490, 11
\reference Villumsen, J.V., Freudling, W., \&
da Costa, L.N., 1996, ApJ, 481, 578

\end{references}
\end{document}